\input epsf.tex

\documentstyle[12pt]{article}

\def\and{{\it\&}}
\def\half{{1\over2}}

\def\quarter{{1\over4}}

\def\gesim{\,{\raise-3pt\hbox{$\sim$}}\!\!\!\!\!{\raise2pt\hbox{$>$}}\,}
\def\lesim{\,{\raise-3pt\hbox{$\sim$}}\!\!\!\!\!{\raise2pt\hbox{$<$}}\,}
\def\boldoverdot{\,{\raise6pt\hbox{\bf.}\!\!\!\!\>}}

\def\ie{{\it i.e.}\ }

\def\ibid{{\it ibid.}\ }
\def\etal{{\it et. al.}\ }
\def\acal{{\cal A}}

\def\lcal{{\cal L}}

\def\ocal{{\cal O}}

\def\diag{\hbox{\diag}}

\def\gev{\hbox{GeV}}
\def\tev{\hbox{TeV}}

\def\m{\hbox{m}}

\def\inbox#1{\vbox{\hrule\hbox{\vrule\kern5pt
     \vbox{\kern5pt#1\kern5pt}\kern5pt\vrule}\hrule}}
\def\sqr#1#2{{\vcenter{\hrule height.#2pt
      \hbox{\vrule width.#2pt height#1pt \kern#1pt
         \vrule width.#2pt}
      \hrule height.#2pt}}}

\def\today{\ifcase\month\or
  January\or February\or March\or April\or May\or June\or
  July\or August\or September\or October\or November\or December\fi
  \space\number\day, \number\year}
\def\pmb#1{\setbox0=\hbox{#1}%
  \kern-.025em\copy0\kern-\wd0
  \kern.05em\copy0\kern-\wd0
  \kern-.025em\raise.0433em\box0 }
\def\lowti#1{_{{\rm #1 }}}
\def\inv#1{{1\over#1}}
\def\su#1{{SU(#1)}}

\def\sumprime_#1{\setbox0=\hbox{$\scriptstyle{#1}$}
  \setbox2=\hbox{$\displaystyle{\sum}$}
  \setbox4=\hbox{${}'\mathsurround=0pt$}
  \dimen0=.5\wd0 \advance\dimen0 by-.5\wd2
  \ifdim\dimen0>0pt
  \ifdim\dimen0>\wd4 \kern\wd4 \else\kern\dimen0\fi\fi
\mathop{{\sum}'}_{\kern-\wd4 #1}}
\font\smalli=cmr8 scaled\magstep1

\def\npb{{\it Nucl. Phys.} {\bf B}}
\def\plb{{\it Phys. Let.} {\bf B}}

\def\prl{{\it Phys. Rev. Let.}}
\def\prd{{\it Phys. Rev.} {\bf D}}
\def\sm{Standard Model}

\begin{document}
\title{CP violation in photon-photon collisions. \hfill {\vskip -.8 in
\hfill {\smalli UCRHEP-T122}}}

\author{\vspace*{1.2em}\\
Jos\'{e} Wudka
\vspace*{0.6em}\\
{\normalsize \sl Department of Physics}\\
{\normalsize \sl University of California, Riverside}\\
{\normalsize \sl Riverside, CA 92521}%
\thanks{Supported in part through funds provided by the
                 Department of Energy and  by the SSC fellowship FCFY9211.}
}

\maketitle


\begin{abstract}
The effective lagrangian parametrization is used to
determine the CP violating effects in $ \gamma \gamma $ collisions.
for the processes studied the effects are found to be very small, the one
exception being scalar production.
\end{abstract}

\section{Introduction}

Probing physics beyond the \sm\ requires high precision experiments
preferably, on quantities whose standard model values are suppressed.
A well known example of this type of observable
is the $ \rho $ parameter, whose values is very close to one due to
the $ \su2_R $ transformation properties of the \sm\
scalar doublet.

A particularly interesting set of processes for which the \sm\
contribution is very much suppressed
consists of those for which CP is
violated. In the \sm\ amplitudes for CP violating processes
are associated with the phase
of the Kobayashi-Maskawa matrix and are extremely small
\cite{Quinn93,Peccei93}. In contrast,
many kinds of new physics generate comparatively
large amounts of CP violation \cite{Gunion92}.
CP violating observables are therefore very good candidates in which to
look for new physics effects.

In this talk I will consider the possibility of observing CP
violating processes
within the gauge-boson and scalar sector of the \sm \cite{Ma93};
part of this work was done in collaboration with J. Gunion and
B. Grzadkowky, a generalization is under investigation.
With only \sm\ interactions these effects are negligible, but this
need not be the case in general. The environment in which I will
study these processes is the photon-photon collider.
Though such a machine will probably be constructed using
back-scattered laser radiation in an $ e^+ e^- $ collider \cite{Ginzburg83},
in this talk I will consider, for clarity and brevity, an
ideal monochromatic
$ \gamma \gamma $ collider in  which both photons can be given any
desired polarization. As I will show, even in this utopian
situation there are great difficulties in observing a clear signal
for some of the processes considered.

The approach which I will follow in this talk is to parametrize the
effects of new physics in a model and process independent way by
using an effective lagrangian \cite{Wudka94}.
This is, by its very nature, a process
and model independent approach which preserves all the successes of
the \sm\ while incorporating new physics in a consistent manner.
I will not describe the formalism in detail here but refer the
reader to \cite{Wudka94}.
Briefly, what is required is to construct all dimension
six operators containing \sm\ fields and respecting the symmetries
of the \sm, which also violate CP. The effective lagrangian consists
of the \sm\ lagrangian plus a linear combination of these operators
with undetermined coefficients. The value of these coefficients
cannot be determined
without further knowledge of the physics underlying the \sm;
nonetheless these couplings can be estimated using consistency
conditions. These estimates are not numerically accurate, nonetheless
they do provide reliable order of magnitude value, which is what is
needed in order to determine the sensitivity of a given
experiment to the scale of new physics.

In this talk I will consider the processes
\begin{equation}
\gamma \gamma \rightarrow
W^+ W^- , \ Z Z , \ H H , \ H . \end{equation}
Some comment on the fermion anti-fermion
final state will be made at the end.

I will assume a ``light'' Higgs \begin{equation} m_H \ll 3~\tev .
\end{equation}

For the processes of interest the relevant dimension six operators are
\cite{Buchmuller86}
\begin{eqnarray}
\ocal_{ \varphi \tilde W } &=& \left( \varphi^\dagger \varphi \right)
W_{ \mu \nu }^I \tilde W_{ \mu \nu }^I \\
\ocal_{ \varphi \tilde B } &=& \left( \varphi^\dagger \varphi \right)
B_{ \mu \nu } \tilde B_{ \mu \nu } \\
\ocal_{ B \tilde W } &=& \left( \varphi^\dagger \tau^I \varphi \right)
B_{ \mu \nu } \tilde W_{ \mu \nu }^I \\
\ocal_{ \tilde W } &=& \epsilon_{ I J K } W_{ \mu \nu }^I
W_{ \mu \nu }^J \tilde W_{ \mu \nu }^K
\end{eqnarray}
So that the lagrangian becomes
\begin{eqnarray}
&& \!\!\!\!\!\!\!\!\!\ \lcal = \lcal_{SM} + \inv{ \Lambda^2 }
\Bigl[
g g' \; \alpha_{ B \tilde W } \ocal_{ B \tilde W }
+ g^3 \alpha_{  \tilde W } \ocal_{  \tilde W } \nonumber
\\ && \quad +  g^2 \alpha_{ \varphi \tilde W } \ocal_{ \varphi \tilde W }
+ g' {}^2 \alpha_{ \varphi \tilde B } \ocal_{ \varphi \tilde B }
\Bigr] .
\end{eqnarray}

The scale $ \Lambda $ determines the limit of applicability of this
parametrization of heavy physics effects: all processes studied using
$ \lcal $ must have energies below $ \Lambda $. Indeed,
the assumption that heavy physics
effects are summarized by a series of effective local operators
can only be true if the energy scale of interest is significantly
smaller than the scale of the heavy physics.
Moreover if we are studying processes whose energies
are such that the underlying physics is apparent, we would
not bother to study their radiative effects in order to re-discover
it. These remarks, though
obvious, are often ignored in the literature.

The coefficients
$ \alpha_i $ will be chosen so that $ \Lambda $ corresponds to the
scale where the heavy physics effects are observed directly.
To estimate them suppose
first that the heavy physics is weakly coupled; in this case
one can verify that
all the  operators $ \ocal $
are generated by loops by the underlying theory. We then expect
\begin{equation} | \alpha_i | \sim{ 1 \over 16 \pi^2 } .  \end{equation}

If the underlying theory is strongly interacting the argument
required to estimate the coefficients $ \alpha_i $ is the
same as the one used in the so-called ``naive dimensional analysis''
\cite{Georgi84}.
The $ \alpha_i $ are in fact running
coupling constants defined by matching conditions at the scale $ \Lambda $,
at which the underlying physics becomes apparent. Then consistency
requires that a change in the renormalization mass $ \mu
\rightarrow c \mu $ with $ c \sim O ( 1 ) $ should not change the
order of magnitude of the $ \alpha_i $. This gives $ | \alpha_{
\tilde W} | \sim 1/ 16
\pi^2 $ and $ | \alpha_{ \varphi \tilde B , \varphi \tilde W ,
B \tilde W } | \sim 1 $.

For a strongly
coupled theory, however, the Higgs mass is expected to receive
large -- $ O(\Lambda)$ -- corrections so that this scenario is
in general inconsistent with the above assumption that the Higgs
is light. The exception occurs when this mass is
protected by a symmetry (such as
supersymmetry). In this case, however, the low energy spectrum of the
models are invariably richer than that of the \sm.
I will therefore assume that a light Higgs is not viable for a natural
strongly coupled heavy theory. In such a situation a different
parametrization, the so called chiral representation,
of the effective lagrangian is required and will not be considered
here due to time limitations. Because of this I will adopt  the
estimates $ | \alpha_i | \sim 1 / 16 \pi^2 $.

\section{Results}

I will then consider an ideal photon collider where the
photons have definite momentum and prefect polarizations.
As mentioned above,
I will not consider the realistic situation where the
photons to be considered are produced by back-scattered laser light.
This is done due to time limitations, and also to avoid complications
which, though quantitatively very important, obscure to a certain
degree the basic problems one has to deal with when trying to uncover
new physics using the processes considered in this talk.

The photons' center of mass momenta are $ k_{ 1 , 2 } = \half \sqrt{s}
( 1 , \pm 1 , 0 , 0 ) $ with polarizations $ \epsilon_{ 1 , 2 }
= \inv{\sqrt{2} } ( 0 , 0 , 1 , \pm \eta ) $, $ | \eta | = 1 $.
In all calculations I will choose $ \eta $ so as to suppress (or in the
optimal case to eliminate) the \sm\ contributions.
When the final particles have the same mass
the final particle momenta are
$ \half \sqrt{s} ( 1 , \pm \beta \cos \theta , \pm \beta \sin \theta , 0 )
$ where $ \beta $ is the velocity of the final particles
and $ \theta $ is the center of mass scattering angle.
With these preliminaries I turn now to the various reactions.

\subsection{$ \gamma \gamma \rightarrow Z Z . $}

To the order we
are working in the effective lagrangian there are no contributions.
The leading terms for this reaction come from dimension eight operators
\cite{Baur93}, and will not be considered further in this talk.

\subsection{$ \gamma \gamma \rightarrow  W^+ W^- . $}

For this process the only contributing operators are $ \ocal_{ \tilde W } $
and $ \ocal_{ B \tilde W } $; if both final $W$ vector bosons are
longitudinal only the second operator contributes. The relevant
diagrams are

\setbox1=\vbox to 2 truein{\epsfxsize=2 truein\epsfbox[0 0 576 792]{f1.ps}}
\centerline{\box1}

\vskip -50pt
\noindent where the solid dot denotes an $ \ocal_{ B \tilde W } $ insertion.

Assuming $ \sqrt{s}  \gg m_W $ allows for the use of the equivalence theorem,
so that the $W$ vector bosons can be replaced by the corresponding
Goldstone particles. For longitudinally
polarized $W$ bosons at these energies in the final state
$ \ocal_{ \tilde W } $ does not contribute, so the final result will
depend on $ \alpha_{ B \tilde W } $ only.

Choosing $ \epsilon_{ 1 2, } = \inv{ \sqrt{2 } }
( 0 , 0 , 1 , \pm i ) $ (\ie $ \eta = i $) yields the amplitude
\begin{equation} \acal( \gamma \gamma \rightarrow  W^+ W^- ) =
{ \alpha \beta_W \over 2 \pi } { t \over \Lambda^2 }
{ 1 \over 1 - \m_W^2 / t } ;  \end{equation}
where $ \beta_W = 16 \pi^2 \alpha_{ B \tilde W } \sim 1 $.
Note that this amplitude is real. The total cross section corresponding
to this amplitude is \begin{equation}
\sigma ( \gamma \gamma \rightarrow  W^+ W^- ) =  { \alpha^2
\beta_W^2 s \over 192 \pi^3 \Lambda^4 } \left[ 1 + O \left(
{ \m_W^2 \over s } \right) \right] . \end{equation}

In the limit of large center of mass energy the \sm\ contribution
vanishes; the amplitude for longitudinally polarized $W$
vector bosons in the final state is \cite{HVeltman93}
\begin{equation}
\acal_{ S M } ( \gamma \gamma \rightarrow  W^+ W^- ) =
- { 8 \pi i \alpha ( 1 - \beta^2 ) \over 1 - \beta^2 \cos^2 \theta } ;
\end{equation}
where $
\beta = \sqrt{ 1 - 4 \m_W
^2 / s } $ (note that it is purely imaginary so it
will not interfere with the $ \ocal_{ B \tilde W } $ contribution). The
corresponding cross section is
\begin{eqnarray}
\sigma_{ S M } &=& { 2 \pi \alpha^2 \over s } ( 1 - \beta^2 ) \times
\nonumber \\
&& \left[ 1 + { 1 - \beta^2 \over 2 \beta } \ln \left( { 1 + \beta
\over 1 - \beta } \right) \right] ;  \end{eqnarray} which
indeed vanishes as $ \beta \rightarrow 1 $:
$ \sigma_{ SM} \simeq 2 \pi  ( 2 \alpha \m_W /s )^2 $
as $ s \rightarrow \infty $.

I will consider the observability of this process from three points of view

\begin{description}
\item[i)] {} Require first
$ \sigma \lowti{ new } \gg \sigma_{ S M } $; this implies
$ s^3/ (1536 \pi^4 \m_W^2 ) \gg \Lambda^4 $. Since we also want
$ \Lambda > \sqrt{s} $ (else the new physics can be probed
directly), this implies $ \sqrt{s} > 31 \tev $; for an accelerator
of this energy scales of order $ 32 \tev $ can be probed.
This result is obtained by assuming
that $ \beta_W = 1 $.

\item[ii)] {} Require $ N \lowti{new } > \sqrt{ N_{ S M } +
N \lowti{ new } } $, where
$ N \lowti{ new } $ is the number of events generated by the
new physics and $ N_{ S M } $ \sm\ events. Using again
$ \Lambda^2 > s $ and $ \beta_W  = 1 $ this condition is equivalent
to the requirement that the luminosity for the machine is greater
than $ 2.4 \times 10^5 $/fb.

\item[iii)] Require that the forward backward asymmetry be greater
than $ 0.1 $ and that there be more than 10  \sm\ events.
This is equivalent to a luminosity above $ 2 \times 10^5 $/fb.
\end{description}

This clearly illustrates the enormous problems one has to deal with:
absurdly large luminosities have to be invoked in order to
detect a signal.

This problem can be traced back to the estimate
$ \beta_W  = 1 $. One might be tempted to relax this condition and
assume, for example, $ \beta_W \sim 16 \pi^2 $ in which case the
required luminosities drop to $ \sim 10 $/fb. Unfortunately such large
values for the coefficients are inconsistent with the whole
approach. In other words, there is no consistent way of generating
such large coefficients from the underlying dynamics without
radically altering the \sm\ itself (for example,
one would then expect $ \rho - 1 = O ( 1 ) $).

\subsection{ $ \gamma \gamma \rightarrow H H .$}

The contributing effective
operators to this process are $ \ocal_{ \varphi \tilde W ,
\varphi \tilde B , B \tilde W } $ appearing in the diagrams

\setbox2=\vbox to 2 truein {\epsfxsize=2 truein\epsfbox[0 0 576 792]{f2.ps}}
\centerline{\box2}

\vskip -50pt

\noindent
where the heavy dot denotes an effective operator insertion.

The \sm\ contributions come from loops such as

\setbox3=\vbox to 2 truein{\epsfxsize=2 truein\epsfbox[0 0 576 792]{f3.ps}}
\centerline{\box3}

\vskip -50pt

\noindent whose evaluation is straightforward.
For simplicity I will consider here
only the expression for the \sm\ generated by a heavy top loop,
this corresponds to the effective operator \cite{Steeger87}
\begin{equation} \ocal \lowti{heavy \ top} = { 35 \over 54 }
{ \alpha G_F \over \sqrt{8} \; \pi }
\left( \half H^2 \right) \left( \quarter F_{ \mu \nu } ^2 \right)
\end{equation}
where $G_F $ is the Fermi constant.
Note that this operator vanishes for the
choice of polarizations
\begin{equation} \epsilon_{ 1 , 2 } = \inv{ \sqrt{ 2 } } ( 0 , 0 , 1 , \pm 1 )
;
\qquad ( \eta = 1 )  \end{equation}
this will be true for all
the \sm\ contributions provided the KM mixing terms are ignored (which
I will do in the following due to the smallness of these effects.

The cross section for this process and for the above choice of polarizations
is \begin{eqnarray} &&
\!\!\! \sigma( \gamma \gamma \rightarrow H H ) =
{ \alpha^2 \beta_H^2 s \over 256 \pi^3 \Lambda^4 } ; \\
&&
\!\!\!\beta_H = 16 \pi^2 \left( \alpha_{ \varphi \tilde W } +
\alpha_{ \varphi \tilde B } - \alpha_{ B \tilde W } \right) .
 \end{eqnarray} Note that $ \beta_H \sim 1 $.

Since the \sm\ contribution vanishes due to the choice of polarization
vectors, the observability of this process is rate dominated.
To estimate the observability of this process I require that 10
events be generated in one year. This corresponds to \begin{equation}
\Lambda \lesim 0.2 \times \sqrt{ \ell \over 100/ \hbox{fb} } \; \tev
 \end{equation} where $ \ell $ denotes the luminosity.

To determine the content of this result recall that, since we are assuming
that the heavy physics is not directly observed, that $ \Lambda
> \sqrt{s} > 2 m_H $. Assuming for example, $ m_H = 250 \gev $ requires
$ \ell > 170 /$fb. On the other hand if $ \Lambda = 1 \tev $  then
$ \ell > 220 /$fb.
For this process the required luminosities are not absurdly
large as in the previous cased, but they are still large requiring
many years' integrated luminosity to observe even a marginal signal.
As before this can be traced to
the consistent estimates of the coefficients in the Lagrangian, if
I had (incorrectly)
taken $ \beta_H \sim 16 \pi^2 $ the required luminosity would
drop by two orders of magnitude.

\subsection{Higgs production.}

The same effective operators considered
above can be used to study the production of single Higgs bosons in
photon colliders. The graph is simply

\setbox4=\vbox to 1.5 truein{\epsfxsize=2 truein\epsfbox[0 0 576 792]{f4.ps}}
\centerline{\box4}

\vskip -50pt

\noindent where the heavy dot denotes an effective operator
vertex.

Note that there is no \sm\ contribution (at tree level),
and that the one loop contributions can be eliminated by the above
choice of polarization vectors. The cross section is then \begin{equation}
\sigma( \gamma \gamma \rightarrow H ) = { \alpha^2 \beta_H^2
m_H \over 32 \sqrt{2} \; \pi G_F \Lambda^4 } \delta( \sqrt{s } - m_H )
 \end{equation} so that, integrating this over the width of the
Higgs and taking \begin{equation} \Gamma_H = { 3 g^2 m_H^3 \over
128 \pi m_W^2 } , \end{equation}
corresponding to a mass above the $WW$ threshold, yields
\begin{eqnarray} \bar \sigma &=&
\int_{ m_H - \Gamma_H/2 }^{ m_H + \Gamma_H/2 }
\sigma( \gamma \gamma \rightarrow H ) { d \sqrt{s} \over \Gamma_H }
\nonumber \\
&=& { 256 \pi^2 \beta_H^2 \over 3 } \left( { m_W s_W \over \Lambda } \right)^4
\inv { m_H ^2 }  \end{eqnarray} so that,
\begin{equation} \bar \sigma \ell = \left( { \ell \over 100 / \hbox{fb} }
\right) \inv{ m_H^2} \left( { 16.2 \over \Lambda } \right)^4 \end{equation}
where $ \Lambda $ and $ m_H $ are measured in \tev.

Taking, for example,
$ m_H = 1 \tev $ and $ \ell = 10 / $fb then values of $ \Lambda $ below
$ 5 \tev $ generate more than ten events.
This is a non-trivial result: a $ 1 \tev $ accelerator can probe scales
five times its energy using this reaction; for lower Higgs mass or higher
luminosities the sensitivity to $ \Lambda $ (as a multiple of
$ \sqrt{s} $) improves.

\subsection{$ f \bar f $ final state}

For the process $ \gamma \gamma \rightarrow f \bar f $ there is
a forbidding zoo of operators that are Cp violating and should be included in
the effective lagrangian describing such processes.
For example,
denoting a left handed fermion doublet by $F$ and a right handed singlet by
$f$,
\begin{eqnarray}
&& i \left( \varphi^\dagger D_\mu \varphi \right)
\left( \bar F \gamma^\mu F \right) - \hbox{ h.c.} ; \nonumber \\ &&
i \left( \varphi^\dagger \sigma_I D_\mu \varphi \right)
\left( \bar F \sigma_I \gamma^\mu F \right) - \hbox{ h.c.} ; \nonumber \\
&& \left( \varphi^\dagger \varphi \right) \left( \bar F f \varphi -
\hbox{ h.c.} \right) ; \nonumber \\ &&
\left( \bar F \sigma^{ \mu \nu } f \right) \varphi B_{ \mu \nu } ;
\end{eqnarray}
\noindent etc. The analysis of such contributions is under way.

\section{Conclusions}

\begin{itemize}
\item  A consistent application of the effective lagrangian
method gives, for most processes considered in this talk,
unobservable rates, even in the optimal situation where the
\sm\ contribution vanishes. This can be traced back to the
consistent estimation of the coefficients of the effective operators.

\item An ad-hoc over-estimation of the coefficients of the
operators can give very nice predictions which might be claimed
to be observable in near future colliders. These results are,
however, completely unreliable being based on an inconsistent model.

\item The best final state here considered is that of single
Higgs production for which  the accelerator becomes a very respectable
probe into the physics underlying the \sm. If there is no Higgs,
or if its mass lies beyond $ 4 \pi v \simeq 3 \tev $, then new interactions
can be expected at this energy. This scenario was not explored in this
talk.

\item The experimentally benign case of a two fermion final
state is currently being studied.
\end{itemize}

\bigskip

The author gratefully acknowledges the help of J. Gunion and B. Grzadkowsy.

\end{document}